\documentclass[12pt,preprint]{aastex}
\usepackage{natbib}

\slugcomment{Will be submitted to {\it The Astrophysical Journal}}

\shorttitle{Tidal Disruption Events}
\shortauthors{Gezari et al.}

\begin{document}

\title{Follow-Up HST/STIS Spectroscopy of Three Candidate Tidal Disruption Events\altaffilmark{1}}

\altaffiltext{1}{Based on observations with the NASA/ESA Hubble Space Telescope obtained
at the Space Telescope Science Institute, which is operated by the
Association of Universities for Research in Astronomy, Incorporated,
under NASA contract NAS5-26555. These observations are associated with
proposal \#9177.}

\author{S. Gezari, J. P. Halpern\altaffilmark{2}}
\affil{Department of Astronomy, Columbia University, 550 West 120th Street, New York, NY 10027-6601}
\email{suvi@astro.columbia.edu}

\altaffiltext{2}{Visiting Astronomer, Kitt Peak National Observatory, 
National Optical Astronomy Observatories, which is operated by AURA, Inc.,
under a cooperative agreement with the National Science Foundation.} 

\author{S. Komossa}
\affil{Max-Planck-Institut f\"ur extraterrestrische Physik, Giessenbachstrasse 1, D-85748 Garching, Germany}

\author{D. Grupe}
\affil{Department of Astronomy, The Ohio State University,
140 West 18th Avenue, Columbus, OH 43210-1173}

\author{K. M. Leighly}
\affil{Department of Physics and Astronomy, The University of Oklahoma,
440 W. Brooks St., Norman, OK 73019}

\begin{abstract}
Large amplitude, high luminosity soft X-ray flares were detected by the {\it ROSAT\/} All-Sky
Survey in several galaxies with no evidence for Seyfert
activity in their ground-based optical spectra.  These flares 
had the properties predicted for a tidal disruption event by a central supermassive
black hole: soft X-ray spectrum, time-scale of months, and large X-ray luminosity
($10^{42} - 10^{44}$ ergs s$^{-1}$).  In order to evaluate the alternative hypothesis
that these flares could have been some form of extreme AGN variability,
we obtained follow-up optical spectroscopy of three of the flaring
galaxies a decade later with the Space Telescope Imaging Spectrograph
(STIS) and a narrow slit to search for
or place stringent limits on the presence of any persistent Seyfert-like
emission in their nuclei.  Two of the galaxies, RX J1624.9+7554 and
RX J1242.6--1119, show no evidence for emission lines or a non-stellar
continuum in their {\it Hubble Space Telescope} ({\it  HST\/})
nuclear spectra, consistent with their ground-based
classification as inactive galaxies.  They remain the most convincing
as hosts of tidal disruption events.  NGC 5905, previously known as a
starburst H~II galaxy due to its strong emission lines, has in its inner
$0.\!^{\prime\prime}1$ a nucleus with narrow emission-line ratios
requiring a Seyfert 2 classification.
This weak Seyfert 2 nucleus in NGC 5905, which was masked by the many surrounding
H~II regions in ground-based spectra, requires a low level of prior
non-stellar photoionization, thus raising some uncertainty about the nature of its X-ray 
flare, which may or may not have been a tidal disruption event.  The absence of
both broad emission lines and nuclear X-ray absorption in NGC 5905
also characterizes it as a ``true'' Seyfert 2 galaxy, yet one that has varied
by more than a factor of 100 in X-rays.
\end{abstract}


\section{Introduction} \label{intsec}

Dormant, supermassive black holes, suspected to be present in the
centers of many normal galaxies, should reveal themselves by a UV/X-ray
flare when they tidally disrupt a star and some fraction of the tidal
debris is accreted.  Tidal disruption flares were proposed by Lidskii \& Ozernoi
(1979) and Rees (1988, 1990) as a probe for supermassive black holes
in the centers of inactive galaxies.  Such an event is very rare in the nucleus
of a galaxy ($< 10^{-4}$ yr$^{-1}$); however, the {\it ROSAT\/} All-Sky Survey (RASS, Voges
et al. 1999) conducted
in 1990--1991 was an ideal experiment to detect these flares since it sampled
hundreds of thousands of galaxies in the soft X-ray band.
{\it ROSAT\/} detected soft X-ray outbursts
from several galaxies with no previous evidence for Seyfert activity [see Komossa (2002)
for a review]:  NGC 5905
(Bade, Komossa, \& Dahlem, 1996), RX J1242.6--1119 (Komossa \& Greiner 1999),
RX J1624.9+7554 (Grupe, Thomas, \& Leighly, 1999), RX J1420+5334 (Greiner et al. 2000),
and a possible fifth candidate, RX J1331--3243 (Reiprich \& Greiner 2001).  NGC 5905,
RX J1624.9+7554, and RX J1420+5334
were detected in the RASS, and then faded beneath the detection limits of the
follow-up pointed observations,
indicating variability by a factor $> 100$ on a time scale of months.
RX J1242.6--1119 was serendipitously observed in the field of view of a pointed observation
18 months after its RASS non-detection,
and was found to have increased in count-rate by a factor $>$ 20.  In addition to the
large amplitude of these flares, they were distinctive due to their extremely soft spectral indices
($2 \leq \Gamma \leq 5$) or blackbody temperatures ($kT_{\rm bb} = 0.05 - 0.1$ keV),
and high luminosities ($10^{42} - 10^{44}$ ergs s$^{-1}$).  From the statistics
of the RASS, Donley et al. (2002) calculated a rate of $\approx 1 \times 10^{-5}$ yr$^{-1}$
for X-ray flares from these non-AGN galaxies.

Table 1 is a summary
of the high-state {\it ROSAT\/} observations for the three galaxies studied in this paper,
including the fitted power-law spectral index or blackbody temperature,
corresponding
0.1--2.4 keV luminosities, absorbing column densities, estimated duration of the flare, and
amplitude of variability.
Previous authors ruled out various outburst scenarios due to their
inconsistencies with the observed properties of the flares.  Stellar sources such
as X-ray binaries and supernovae rarely reach X-ray luminosities exceeding 10$^{40}$ ergs s$^{-1}$.
Gamma-ray burst afterglows could reach high enough peak luminosities, however, they have harder spectra and shorter durations than these flares,
and are accompanied by bright optical afterglows.  Gravitational lensing is another
candidate for a high luminosity flare, but such an event would cause amplification at all
wavelengths, not just in soft X-rays.   The remaining outburst mechanisms that are able to
produce the observed large flare luminosities require a central supermassive black hole.  Soft
X-ray variability
due to varying obscuration or warm absorption of a Seyfert nucleus in NGC 5905 was examined in detail
by Komossa \& Bade (1999).  Since no Seyfert-like narrow emission lines or broad Balmer lines
were detected in ground-based optical spectra, their warm absorber model included dust
to explain the absence of such optical emission.  The dusty warm absorber
did not allow an acceptable fit to the high-state and low-state {\it ROSAT\/} spectra without
fine-tuning of the dust composition.  With an obscured Seyfert nucleus as an
unlikely fit, and no evidence for persistent Seyfert activity at any wavelength, an
accretion disk instability was also disfavored as the cause for the
soft X-ray outburst.  A tidal disruption event by an otherwise quiescent central
supermassive black hole arose as the most likely cause for the flares in these apparently
inactive galaxies.

Throughout this paper, calculated and quoted luminosities are for $H_0 = 75$ km~s$^{-1}$~Mpc$^{-1}$ and $q_{0} = 0.5$.

\section{Theory of Tidal Disruption Events} \label{theory}

When a star of mass $m_*$ and radius $R_*$
is in a radial ``loss cone'' orbit that brings it within the tidal
radius of a central supermassive black hole,
$R_T = (M_{\rm BH}/m_*)^{1/3}R_*$, the tidal forces on the star are
stronger than its self-gravity, and the star will be torn apart.  For a solar type star,
$R_T = R_{\sun}(M_{\rm BH}/M_{\sun})^{1/3} = 1.1 R_{\rm Sch}M_8^{-2/3}$,
where $M_8 = M_{\rm BH}/10^8 M_{\sun}$. If
$M_{\rm BH} > 1.1 \times 10^8 M_{\sun}$, then the tidal radius is less than the Schwarzschild radius,
and solar-type stars will not be disrupted, but will be swallowed whole by the black hole.  For this
mass range, red giants could be stripped of their outer layers, and solar-type stars could be
disrupted if the black hole is spinning.  The rate
at which stars are disrupted by a central black hole depends on the mass of the black hole,
the stellar density and velocities
in the galactic nucleus, and the rate at which radial loss cone orbits are depleted.
For typical nuclear stellar densities and velocities in the centers
of galaxies, this results in
a tidal disruption rate of $10^{-6} - 10^{-4 }$ yr$^{-1}$ (Magorrian \& Tremaine 1999;
Syer \& Ulmer 1999).
When a star is disrupted,
half of the tidal debris will be ejected on hyperbolic
orbits, half will return to pericenter and circularize, and a fraction of that bound
debris will be accreted by the central black hole (Rees 1988).  The return
rate of this bound material on orbits that intersect at pericenter is what determines
the time scales of the onset and decay of the luminosity resulting from a tidal disruption
event.  The observable flare would begin with accretion of the most tightly bound material
as it returns to pericenter and circularizes.  As summarized by Ulmer (1999), the first
return would occur at a time $t_0$ after the disruption, where $t_0 \sim 1.1 M_8^{1/2}$ yr. 
The maximum return rate, according to numerical simulations (Evans \& Kochanek 1989) is
$\dot M_{\rm max} \sim 0.14\,M_{8}^{1/2}\,M_{\sun}$ yr$^{-1}$, and occurs at a time $t \sim 1.5\,t_0$.
At late times, material returns at the declining rate $\dot M(t) = 0.3\,M_8\, (t/t_{0})^{-5/3} M_{\sun}$
yr$^{-1}$, which is an important factor that controls the decay of the flare over the next few
years.  For $M_{\rm BH} < 10^{7}\ M_{\sun}$, the dynamical time for the tidal debris is short relative
to the return time, so the material is accreted quickly, resulting in a super-Eddington 
flare ($L_{\rm flare} > \epsilon M_{\rm Edd} c^{2} > 1.3 \times 10^{46} M_8$ ergs s$^{-1}$)
(Ulmer 1999).  For $M_{\rm BH} > 2 \times 10^7 M_{\sun}$, stellar debris takes a longer time to fall back than
the rate at which it can circularize and radiate, thus accretion probably proceeds through a 
thin disk, with a luminosity always less than the Eddington limit.  For a super-Eddington infall rate, the 
duration of the accretion flare is $t_{\rm flare} \sim 0.76\,M_7^{-2/5}$ yr, and the
spectrum of the flare will be characterized by the 
blackbody temperature of a thick disk or spherical envelope at the tidal radius,
$T_{\rm eff} \approx (L_{\rm Edd}/4\pi\sigma R_T^2)^{1/4} = 3.0 \times 10^5 M_{7}^{1/12}$ K  (Ulmer 1999).
Thus, for low-mass central black holes ($M_{\rm BH} < 10^7 M_{\sun}$), tidal disruption
theory predicts luminous flares of up to $10^{46}$ ergs s$^{-1}$, 
peaking in the UV/X-ray domain, with time scales on the order of months.

The three flaring galaxies studied in this paper had X-ray outbursts with properties predicted
for a tidal disruption event:  a high-state spectrum well-fitted by a
soft blackbody and a flare duration on the order of months. The galaxy
with the best sampled X-ray light curve, NGC 5905, shows a fading of the flare luminosity close to the
predicted accretion rate $\dot M \propto t^{-5/3}$ (Komossa \& Bade 1999), which Li, Narayan, \&
Menou (2002) regard as strong evidence that its flare was a tidal disruption event.
Although the tidal disruption hypothesis is appealing, there remains the possibility 
that low-luminosity Seyfert nuclei
could reside in these galaxies, undetected in their low-states by ground-based 
optical observations.  In order to establish beyond a reasonable doubt whether
these flares were in fact tidal disruption events, as opposed to some
other form of 
extreme AGN variability, we obtained follow-up
{\it HST\/}/STIS spectroscopy through narrow slits, which are
up to a factor of 100 more sensitive to nuclear activity than previously obtained
ground-based spectra, to search for persistent Seyfert activity in the centers
of these galaxies.

\section{Observations and Basic Results} \label{obssec}

Long-slit spectra centered on the nuclei of RX J1242.6--1119A (the larger of a pair
of galaxies at $z = 0.050$), RX J1624.9+7554 (a barred spiral
at $z = 0.064$),
and NGC 5905 (a large barred spiral at $z = 0.011$), were obtained a 
decade after their flares
in Cycle 10 of {\it HST\/} with the STIS/CCD and either a $0.\!^{\prime\prime}2$ or a
$0.\!^{\prime\prime}1$ slit.   In each case, pairs of
acquisition images were first
obtained that were used to center the slit precisely on the nucleus of the galaxy.
The STIS 2D spectra
were reduced using the data calibration pipeline CALSTIS.  After 
sky
subtraction and bad-pixel removal, a 1D spectrum was rectified using a width in the spatial
direction of $0.\!^{\prime\prime}36$, corresponding to 93.5\% of the total energy
in the point-spread-function (PSF).  
For comparison with the STIS spectra, ground-based spectra of comparable
spectral resolution were obtained for each galaxy on the KPNO 2.1m
and MDM 2.4m telescopes, with slits of width $1.\!^{\prime\prime}5 -
1.\!^{\prime\prime}9$.  Table 2 is a
log of STIS and ground-based observations.
The STIS spectra of RX J1242.6--1119A and RX J1624.9+7554, shown in Figures 1 and 2, 
have stellar absorption features that are very similar to their ground-based
optical spectra.  Although the STIS spectra were
obtained through a narrow slit, which explains the factor of $\sim 5-10$ weaker
continuum levels compared to the ground-based spectra, there appears to be no difference in their spectral features or continuum shape.
Some weak extended [N~II] emission is seen in the KPNO 2.1m spectrum of RX J1624.9+7554,
which was also detected in the ground-based spectrum of Grupe et al. (1999), and is most 
likely from the photoionized or shocked diffuse interstellar medium 
near the nucleus.
Thus, there is no evidence for nuclear emission lines
or a non-stellar continuum indicative of an AGN in
RX J1242.6--1119A or RX J1624.9+7554.  These two galaxies
remain the most convincing as hosts of tidal disruption
events, and will not be discussed further here.
The remainder of this paper is concerned with evaluating the
more problematic case of NGC 5905.

The STIS spectrum of NGC 5905, shown in Figures 3 and 4, 
contains Balmer and forbidden emission lines, including H$\alpha$, H$\beta$, 
[O~III], [O~II], and [N~II].  The ground-based spectra differ in both continuum level
and relative narrow emission-line strengths.  Most dramatic are the differences
in the strength of [O~III]$\lambda\lambda$4959,5007 relative to H$\beta$, and [N~II]$\lambda\lambda$6548,6583 relative
to H$\alpha$.  The forbidden lines are strongest relative to the Balmer lines
in the STIS spectrum of the nucleus, whereas the
STIS spectra offset to the east and west of the nucleus
(Figure 4) have emission-line ratios similar to the ground-based spectrum.
The blue-shifted H$\alpha$ and [N~II] lines in the $+1.\!^{\prime\prime}4$ offset spectrum and
the red-shifted lines in the $-1.\!^{\prime\prime}2$ offset spectrum
align with the blue and red peaks
of the double-peaked
H$\alpha$ and [N~II] lines in the STIS nuclear spectrum, which we interpret
as rotation.

Figure 5 shows an H$\alpha$
rotation curve of the central $2^{\prime\prime}$ of NGC 5905 measured from the 2D STIS spectrum, in which the slit was oriented at position angle $99.\!^{\circ}4$,
nearly perpendicular to the bar.  The red-shifted and blue-shifted
components of the double-peaked nuclear H$\alpha$ line are seen to
match the velocities of the outer rotation curve.  This can be interpreted
as a nuclear gas disk rotating in the same plane as the outer gas disk.
The observed radial velocity semi-amplitude, $v_r = 114$ km~s$^{-1}$,
is assumed to be the line-of-sight
component of the circular rotation velocity $v$, where
$$v_r\, =\, {v\, {\rm cos}\,i\ {\rm sin}\, i \over ({\rm cos}^2\,i + {\rm tan}^2\,\phi)^{1/2}}\ ,$$
$i$ is the inclination angle of the rotation axis relative to the line of sight, and
$\phi$ is the angle between the spectrograph slit and the major axis of the galaxy.
Similarly, the projected distance $d$ along the slit is related to the physical radius $r$
in the galaxy via
$$d =\, r\, {\rm cos}\,i \left(1+{\rm tan}^2\,\phi \over {\rm cos}^2\,i + {\rm tan}^2\,\phi\right)^{1 \over 2}.$$
From the UGC catalog (Nilsson 1973) and other sources (van Moorsel 1982; Ma 2001),
we adopt $i = 40^{\circ}$, and we determine that
$\phi \approx 35.\!^{\circ}6$.  For this set of parameters, we derive $v = 243$ km~s$^{-1}$,
which is in excellent agreement with the maximum H~I rotation velocity of 245 km~s$^{-1}$
measured by van Moorsel (1982).
Accordingly, the derived enclosed mass $M_{\rm enc} = r\,v^2/G \leq 1.7 \times 10^8 M_{\sun}$
within an aperture $d < 11\,{\rm pc} = 0.\!^{\prime\prime}05$
(half the width of the slit).  This mass estimate is
consistent with $M_{\rm BH} \lesssim 1 \times 10^8 M_{\sun}$, and therefore in the
appropriate mass range for causing the tidal disruption of a solar mass star.

Emission-line strengths in the STIS spectra of the nucleus of NGC5905
were measured using a wavelength interval of 27.4 \AA\ for the low-resolution
G430L spectrum, and 11.1 \AA\ for the medium-resolution G750M
spectrum (in order to include all the flux
from the double-peaked profile of the lines).  The neighboring continuum was
estimated as the median value in an interval of the same width on either side of
the line, with the standard deviation in the continuum
interval estimated as the $1\sigma$ error.  The emission lines in the high signal-to-noise ground based-spectrum,
shown in Figure 4, were measured using Gaussian fits to the lines.  Table 3 lists the optical
line fluxes relative to H$\alpha$ for both the STIS and ground-based spectra of NGC 5905.  The 
nuclear spectrum is significantly reddened, as indicated by the
ratio H$\alpha$/H$\beta = 11.8$.  Also listed in Table 3 are the
STIS line fluxes corrected for reddening using the extinction curve
of Cardelli, Clayton, \& Mathis (1989) and visual extinction $A_V = 4.26$ magnitudes,
assuming intrinsic H$\alpha$/H$\beta = 3.1$ as is appropriate for Seyfert narrow-line
regions (Halpern \& Steiner 1983).  This value of $A_V$ should be considered
an upper limit, as it does not take into account probable underestimation
of the flux of H$\beta$ emission due to its superposition on stellar absorption
from which it cannot be separated.  When the double-peaked [N~II] and H$\alpha$ lines 
in the STIS G750M nuclear spectrum are fitted with two Gaussians, the width of each Gaussian, corrected for
instrumental resolution, is 130 km s$^{-1}$.  The single-peaked [N~II] and H$\alpha$ lines in the 
offset spectra are narrower, with an intrinsic width of 80 km s$^{-1}$.  This difference
indicates that there is a source of broadening in the nucleus in addition to rotation.  
The intrinsic width of the
H$\beta$ and [O~III] lines in the STIS G430L nuclear spectrum is 400 km s$^{-1}$, broader
than the lines in the G750M nuclear spectrum, probably because the spectral
resolution of the G430L spectrum is inadequate to resolve the double-peaked structure of the lines.  

\section{Interpretation} \label{discsec}

\subsection{Ground-based Versus {\it HST\/} Post-flare Optical Spectra of NGC 5905} \label {lr}

The post-flare ground-based and {\it HST\/} STIS spectra
of NGC 5905 differ dramatically.  The stellar continuum flux in the STIS spectrum is lower than the
KPNO spectrum by two orders of magnitude, and the narrow emission line ratios in
the STIS spectrum indicate a higher ionization state for the line emitting gas. 
The narrow emission line ratios measured from the KPNO spectrum identify NGC 5905
as a H~II/starburst galaxy, consistent with the interpretation of previous ground-based spectra (Ho et al. 1997; Komossa \& Bade 1999).
The STIS spectrum of the nucleus of NGC 5905, on the other hand, has narrow emission-line 
ratios that place it in the Seyfert region of the diagnostic diagrams of
Veilleux \& Osterbrock (1987, see Figure 6).
These differences in continuum flux and narrow emission-line ratios can be understood as a result
of the narrower $0.\!^{\prime\prime}1$ slit used for the STIS
spectrum as compared to the $1.\!^{\prime\prime}9$ slit of the KPNO spectrum,
appropriate for ground-based seeing.  The narrow slit effectively 
reduces the contribution of stellar light and circumnuclear H~II region emission in the STIS
spectrum, making it more sensitive to flux originating from the nucleus itself.  The 
narrow-line emission in the STIS spectrum is spatially compact compared to
the stellar continuum (shown in Figure 7), 
indicating that the STIS 
has detected emission from a distinct source in the nucleus, namely,
a low-luminosity Seyfert 2 nucleus that was previously masked by H~II regions in the
ground-based spectra.

\subsection{Search for Non-Stellar Continuum \& Broad Emission Lines}

The post-flare STIS spectrum of NGC 5905 contains no broad Balmer line
emission and no direct evidence for a non-stellar continuum.  The Mg I 
$b$ $\lambda$5175 and G-Band $\lambda$4300 stellar absorption features are 
detected in the STIS spectrum (shown with dotted lines in Figure 3), 
with equivalent widths consistent with the absorption features in the 
KPNO spectrum within $2\sigma$, where $1\sigma$ $\sim$ 1.2 \AA.  The $1\sigma$ 
error includes the statistical error estimated from the neighboring 
continuum, and the systematic error from choosing the continuum level.  
The error in the equivalent widths places an upper limit on the percent 
contribution by a featureless continuum of 60\%.  This featureless 
continuum could be a power-law continuum from the Seyfert nucleus, or 
due to young stars.  The Ca II H \& K $\lambda\lambda$3968,3933 features 
and 4000 \AA\ break are not obvious in the STIS spectrum, however their 
presence cannot be ruled out at this signal-to-noise level.  It should be 
noted that the depth of the 4000 \AA\ break is sensitive to the stellar 
population, as well as to the presence of an AGN continuum.  The spatial 
profile of the acquisition image, shown in Figure 8, also does not 
indicate an AGN-like unresolved point source in the nucleus.  Its radial 
profile is only slightly more concentrated
than the confirmed inactive galaxies RX J1242.6--1119A and RX J1624.9+7554.

While these optical properties of NGC 5905 might be
consistent with a Seyfert 2 nucleus having an obscured broad-line region
and non-stellar continuum, the lack of X-ray absorption during the soft
X-ray flare is a strong indicator that the direct line of sight to the nucleus
is unobscured.  Thus, the lack of broad emission lines is either
intrinsic, or is evidence that the nucleus is in a quiescent state.
The X-ray flare will have
illuminated any gas around it with different light travel time delays.  The observer
will see reprocessed emission from an isodelay paraboloid, 
with the source of the flare at the focus.  A point at a distance $r$
from the flare and at
angle $\theta$ from the line from the source to the observer
will be illuminated with time delay $\tau = (1-{\rm cos}\,\theta)\,r/c$.  Gas
in the line of sight ($\theta = 0^{\circ}$) will be illuminated with no time
delay, and gas behind the nucleus ($\theta = 180^{\circ}$), will
be observed to be illuminated with the maximum time delay of
$2\,r/c$.  Given that the STIS spectrum was obtained 11.4 years after the
X-ray flare was observed,
the smallest radial distance that the flare could be probing in the
STIS spectrum is 1.7 pc, or 5.7 lyr.  This is orders of magnitude larger
than the scale of the broad-line region in Seyfert galaxies, which is typically
on the order of light days or weeks (e.g., Wandel, Peterson, \& Malkan 1999).  
Therefore, we do not expect to see flare-excited broad-line emission in the STIS
spectrum.  In addition, since the high density typical of the broad-line region
corresponds to rapid recombination times, fossil broad-line emission
would not be expected either.

This result can be compared to observations of IC~3599, a {\it ROSAT\/} flaring galaxy that 
was observed optically only 5 months after its X-ray flare (Brandt, Pounds, \& Fink
1995).  In that case, strong Balmer emission lines and highly ionized [Fe VII] and
[Fe X] lines 
characteristic of a Seyfert galaxy were detected.   In the next spectrum obtained
9 months later, the highly ionized Fe lines had disappeared (Grupe et al. 1995).  Henceforth,
the optical spectrum of IC 3599 maintained the characteristics of a Seyfert galaxy
(Grupe et al. 1995) of type 1.9 (Komossa \& Bade 1999).  Thus, the first post-flare
spectrum had detected Fe emission from gas within 0.1 pc of the nucleus, ionized by the 
variable continuum.  The opportunity to 
detect line emission from gas ionized by the flare very close to the nucleus in NGC 5905 
was lost because the first post-flare optical spectrum 
was not obtained until 6 years after the event (Komossa \& Bade 1999).

\subsection{Narrow Emission Lines:  Low-luminosity Seyfert 2 or Fossil Nebula?} \label {so}

Even in the absence of other evidence for Seyfert activity,
the narrow emission-line ratios in the STIS spectrum
of NGC 5905 require a non-stellar ionizing continuum.  Two
possibilities for the excitation of the line emission can be considered: narrow-line
clouds ionized by a persistent,
low-luminosity Seyfert nucleus, or the fossil nebula of the
ionizing soft X-ray flare.  If the emission lines are powered by
a Seyfert nucleus, then the dereddened H$\alpha$ flux can be used to
predict the time averaged
soft X-ray luminosity of the nucleus using the correlation of soft X-ray
luminosity with H$\alpha$ luminosity derived from observations of low-luminosity AGN ($L_X < 10^{42}$ 
ergs s$^{-1}$) by Halderson et al. (2001), who find $L_X = 7\,L_{{\rm H}\alpha}$ on average.
The predicted soft X-ray luminosity of NGC 5905 is 
$\sim 4 \times 10^{40}$ ergs s$^{-1}$, consistent with the
lowest value
observed by the {\it ROSAT\/} PSPC ($L_X \approx 2 \times 10^{40}$ ergs s$^{-1}$), but
the latter detection might
also be partly associated with its nuclear starburst (Bade, Komossa, \& Dahlem, 1996).

If the narrow emission lines are in fact powered by a Seyfert nucleus, then
photoionization modeling can constrain the density and covering fraction $f_c$
of the line-emitting gas.  Using the photoionization code of Halpern \& Steiner (1983),
and similar works by Ferland (1981) and Ferland \& Netzer (1983), we come to the following
conclusions.  The H$\beta$ luminosity as an indicator
of the rate of absorption of ionizing photons yields $f_c\,Q({\rm H}) \approx 5.4 \times 10^{51}$ s$^{-1}$.
Employing the ionization parameter $U\, =\, Q({\rm H})/(4\pi\,r^2\,n_{\rm H}\,c)$, which must
be $\sim 10^{-3}$ for the excitation
level indicated by the observed [O~III]/H$\beta$ flux ratio,
we find $f_c\,n_{\rm H}r^2 \approx 1.5 \times 10^{43}$ cm$^{-1}$.  Since $r \leq 11$~pc,
we require $f_c\,n_{\rm H} \geq 1.3 \times 10^4$ cm$^{-3}$.  
The luminosity of the collisionally excited [O~III]
line also constrains the density and covering fraction of the line emitting gas.
Assuming a plane-parallel geometry at a distance $r$ from the nucleus,
a thickness $\Delta r < r$ for [O~III] zone of the cloud or filament, and a covering fraction $f_c$, the luminosity of [O~III] is:
$$L({\rm [O~III]}) = h\nu_{\rm [O~III]}4\pi r^2\Delta r\,n_{\rm O^{++}}\,n_e\,q_{12}\,f_c\ 
{\rm ergs\ s}^{-1} $$
For a collisional excitation rate $q_{12} = 2.6 \times 10^{-9}$
cm$^{3}$~s$^{-1}$ at $T = 15,000$~K, n$_{\rm O^{++}} = 
10^{-4}\,n_e$, and a dereddened [O~III] luminosity of $1.1 \times 10^{40}$ ergs~s$^{-1}$, this implies an emission measure
$r^2\Delta r\,n_e^2\,f_c = 8.5 \times 10^{62}$ cm$^{-3}$.  Models indicate that
$\Delta r\,n_e$ for the [O~III] zone is $\approx 1 \times 10^{20}$ cm$^{-2}$ for $U \approx 10^{-3}$,
so with $r \leq 11$~pc (half the width of the slit), we find $n_e\,f_c \geq 7300$.
Another constraint on $n_e$ comes from the [S~II] density-sensitive line ratio, which
indicates higher density in the nucleus than in the H~II disk.  Allowing for
the large error bars on the fluxes of these weak lines, we find
$0.3 < F(\lambda 6716)/F(\lambda 6731) < 1 $ in the nucleus, which implies $n_e \geq 500$
(i.e., consistent with the high-density limit for this diagnostic).  Thus, photoionized
clouds of $n_{\rm H} \sim 10^4$ and large convering fraction are consistent with
the STIS spectrum of the nucleus.  To the extent that we have overestimated the extinction,
$n_{\rm H}$ and/or $f_c$ might be smaller.

Although a persistent Seyfert nucleus cannot be ruled out as the source of the high excitation narrow emission lines,
could the X-ray flare have produced these lines that we observe 11 years later?
Using the blackbody fit to the {\it ROSAT\/} spectrum (which is
more reasonable than a steep power-law, which would extrapolate to
a bright optical flare that was not observed)  the number of
photons available to ionize hydrogen at the time of the flare can be estimated
as those between the Lyman limit $\nu_0$ and 100 eV,
$$Q({\rm H}) = \int_{\nu_{0}}^{100\,{\rm eV}} \frac{L_{\nu}}{h\nu} d\nu = 
4\pi R_{\rm bb}^{2}\,\frac{2\pi}{c^{2}}\left(\frac{kT}{h}\right)^{3}\int \frac{x^{2}}{e^{x}-1} dx = 1.5 \times 10^{51}\ {\rm photons\ s}^{-1},$$
\noindent where R$_{\rm bb} = 9.2 \times 10^{10}$ cm [corresponding to a $0.1-2.4$ keV X-ray luminosity of
$9.2 \times 10^{41}$ ergs s$^{-1}$ from Bade, Komossa \& Dahlem (1996)]
and $T = 6.5 \times 10^5$ K.
The predicted peak luminosities of H$\beta\,\lambda4861$ and He~II $\lambda4686$
are
$$L = f_c h\nu\frac{\alpha_{\nu}^{eff}}{\alpha_{B}}\int_{\nu_{0}}^{100\,{\rm eV}} \frac{L_{\nu}}{h\nu} d\nu\ {\rm ergs\ s}^{-1},$$
\noindent yielding $L = 5.4 \times 10^{38} f_c$ ergs s$^{-1}$ for H$\beta$ and $1.6 \times 10^{39} f_c$ ergs s$^{-1}$ for He~II, with $\nu_0$(He) = $4\,\nu_0$(H).  The dereddened STIS 
luminosity of H$\beta$ is $1.9 \times 10^{39}$ ergs s$^{-1}$, 2.5 times greater than the peak luminosity powered
by the flare, while He~II is not detected with more than 2$\sigma$
significance in the STIS spectrum, and can be assigned an upper 
limit of $6.5 \times 10^{38}$ ergs s$^{-1}$.  The non-detection of the He II line in the STIS spectrum,
which would be expected to be even stronger than the H$\beta$ line for gas ionized by the flare,
is strong evidence that its emission lines are not powered by the flare, or if they are, that they must have decayed significantly. 

The decay time
for each emission line depends inversely on the density of the gas.
Using the time-dependent 
photoionization model results of Binette \& Robinson (1987), narrow-line emitting gas should continue to emit 
after the ionizing source (i.e., X-ray flare) has turned off.  Although the [O~III] line 
responds rapidly to the change in ionizing flux, and decays almost instantaneously due to efficient 
recombination by charge transfer, the He~II and H$\beta$ lines will persist after
the flare.  Neglecting for the moment light travel time delays,
the decay time for [O~III] as a function of density is $\tau = 2000\,n_{\rm H}^{-1}$ yr, 
thus for $n_{\rm H} > 180$ cm$^{-3}$ 
the decay time is less than 11 years, and any [O~III] emission excited
by the flare will have decayed by now.  The He~II line has a decay time, $\tau = 23,500\,n_{\rm e}^{-1}$ yr,
such that for $n_e < 2350$ cm$^{-3}$, there should be fossil He~II emission still detected
in the post-flare spectrum 11 years later.  Given the typical range of densities for narrow-line
regions, if the line emitting gas were powered by a soft X-ray flare, the
post-flare spectrum observed 11 years later should contain He~II emission,
but most if not all of 
the [O~III] line emission will have decayed away. 
Given that [O~III] is the stronger emission line compared to 
H$\beta$ and He~II in the STIS spectrum, the narrow emission lines do not have flux ratios
diagnostic of such a fossil nebula, i.e. [O~III]/He~II $\leq$ 1 (Binette \& Robinson 1987).  

If we now allow for light travel time delays, we might hypothesize that the reason
the narrow-line emission does not have line ratios characteristic of a fossil nebula 
is that we are seeing only clouds recently illuminated by the flare, and that
differential rates of photoionization of various species could account for the
strong [O~III] emission relative to H$\beta$ and He~II.  The photoionization rate for hydrogenic ions is
$$\int_{\nu_{0}}^{100\,{\rm eV}} \frac{F_{\nu}}{h\nu}\,\sigma_{\nu}(Z)\,d\nu\ \ {\rm s}^{-1}$$
\noindent
where $F_{\nu} = L_{\nu}/4\pi r^2$ and $\sigma_{\nu}(Z)$
is the photoionization cross section as a function of atomic number $Z$.
The resulting photoionization time scales at $r = 11$ pc (the half width
of the STIS slit) are 115 days for H and 96 days for He.
Similarly, the collisional
excitation rate of [O~III] at $15,000$ K is $2.6 \times 10^{-9}$ cm$^3$ s$^{-1}$,
corresponding to a mean excitation time of $9\,(n_e/500)^{-1}$ days.
Since the
excitation times for all of these emission lines are very short
relative to the light travel time between different parts of the narrow-line
region, we would not expect to observe effects of photoionization delays.

If light-travel time delays are invoked to explain the continued presence of
[O~III] emission in the STIS post-flare spectrum of NGC 5905, then this
argument cannot simultaneously account for the absence of
He~II emission, which would be strong in recently ionized clouds.
Thus, the line ratios in the STIS spectrum are not consistent with
clouds photoionized by a soft X-ray flare, either near the nucleus and now
decaying, or far from the nucleus and just being ionized.
Therefore, the narrow line emission observed in NGC 5905 requires there
to be a long-term source of ionization, i.e., a persistent Seyfert nucleus.

\subsection{Geometry of the Line-Emitting Gas} \label {go}

If NGC 5905 in fact contains a Seyfert 2 nucleus, then according to the unified 
model it should be surrounded by a dusty torus that obscures emission in the 
line-of-sight from rapidly orbiting clouds in a broad-line region close to the central 
accreting black hole (Antonucci 1993). If this torus obscures
the nucleus, then how could a soft X-ray flare 
have been observed from the vicinity of the black hole through such a high column density of X-ray 
absorbing gas?  The column density derived from the {\it ROSAT\/} spectrum is close to the Galactic value,
$1.5 \times 10^{20}$ cm$^{-2}$, while the Balmer decrement in the STIS spectrum corresponds to a
column density of $8 \times 10^{21}$ cm$^{-2}$.  This discrepancy might be addressed by
invoking a higher than Galactic dust-to-gas ratio for the line emitting gas, or a gross
error in the H$\alpha$/H$\beta$ measurement in the STIS spectrum due primarily to stellar 
absorption in those lines.  However, neither effect is likely to be large 
enough to account for the factor of $\approx 50$ difference in
column density.  A geometrical argument, on the other hand, could easily explain the 
observed optical extinction and lack of X-ray absorption
if dusty line-emitting gas is distributed in a disk or torus in the plane
of the galaxy and out of our line of sight to the nucleus.  The double-peaked
line profiles that match the rotation curve of the Galaxy also support this interpretation.

If the nucleus of NGC 5905 is not obscured by dust in the line-of-sight,
then why does the STIS spectrum show only Seyfert 2 features? 
Halderson et al. (2001) found that Seyfert 2 galaxies in their sample of low-luminosity
AGNs did not
have systematically higher column density than Seyfert 1 galaxies in fits
to {\it ROSAT\/} spectra.  While this is contrary to the prediction of the unified
model, they also attributed the apparent similarity in column density to the 
contamination by extended soft X-ray emission observed in both types of low-luminosity
Seyfert galaxies.  Ho et al. (2001) also found small column densities, $N_{\rm H} < 10^{22}$ cm$^{-3}$,
for most of the low-luminosity AGNs in their survey,
including Seyfert 2s, suggesting that the unified model for Seyfert galaxies might not
apply at low luminosities.  An alternative interpretation of Seyfert 2 galaxies with small measured 
line-of-sight gas column densities is that they are ``true'' Seyfert 2s, with 
no broad line region, and any extinction is due to large-scale structures such as dust-lanes
or circumnuclear starbursts.  Panessa \& Bassani (2002) find that unobscured Seyfert 2's, with observed 
X-ray column densities $N_{\rm H} < 10^{22}$ cm$^{-2}$, are biased towards lower luminosities.  
There is also mounting observational evidence that correlates a lower luminosity with a smaller broad 
line region (Wandel et al. 1999; Kaspi 2000; Collin \& Hur\'e 2001).  
This model can be extended to low enough luminosities that no broad line region exists (Tran 2001).
An intrinsically weak Seyfert nucleus can also account for the non-detection of a featureless
continuum in the STIS spectrum, and the lack of optical variability (within $\sim 0.2$ mag) 
during the flare as reported by Komossa \& Bade (1999) from photographic plates.  
Thus, NGC 5905 may have an intrinsically weak non-stellar continuum,
no broad line region, and only narrow-line clouds available to be ionized.  A final
possibility to consider is that the {\it HST}/STIS observations could have occurred 
during a low-state of the Seyfert nucleus, 
when the ionizing continuum and broad line region emission are temporarily ``off''.

\section{Conclusions} \label{dissec}

RX J1242.6--1119A and RXJ1624.9+7554 are confirmed to be inactive galaxies
from the lack of broad or narrow emission lines, or a non-stellar continuum in their
narrow-slit {\it HST\/}/STIS spectra.  The most likely explanation for large amplitude
X-ray flares from the nuclei of inactive galaxies is a tidal disruption of a star
by an otherwise dormant central supermassive black hole.

An important question is raised in the case of NGC 5905:
whether its X-ray flare was the result of
large amplitude variability of its low-luminosity Seyfert nucleus, 
or a tidal disruption event.
Both scenarios involve a dramatic increase in the rate of accretion 
onto the central black hole, either by an instability in a
preexisting accretion
disk, or by the formation of a debris disk from the tidally disrupted star.
We use emission-line diagnostics
to show that a persistent Seyfert nucleus
in NCG 5905 is required to account for its nuclear emission lines,
which we resolved from the
surrounding H~II regions using {\it HST\/}/STIS spectroscopy.
While not ruling out the possibility that the
soft X-ray flare was a tidal disruption event, this prior activity does raise some
uncertainty about such an interpretation.
The nuclear emission-line luminosities correspond to
an average X-ray luminosity of
$\approx 4 \times 10^{40}$ ergs s$^{-1}$, which is similar to the
lowest value seen by {\it ROSAT\/} after the flare.
Further long-term monitoring in the X-ray and
optical is necessary to see whether flaring behavior in NGC 5905
recurs, which would indicate variability intrinsic to the Seyfert nucleus,
or whether its X-ray flare was a one-time event consistent with the
expected frequency of tidal disruption.  If the flare
was not a tidal disruption event, it is one of the most extreme cases 
of X-ray variability observed sofar among AGN.  A separation of nuclear and
starburst X-ray sources in NGC 5905 can be accomplished by {\it Chandra},
which could then establish the minimum X-ray luminosity of the nucleus
by repeated observations.    

It is important to find unambiguous indicators of tidal disruption events.
If they have properties similar to the candidates studied here, then
future UV and X-ray sky surveys will have the potential to detect
hundreds of tidal disruption events, making it possible to study the masses of black
holes as a function of galaxy type and size, unbiased by the minority of
galaxies that are AGNs, and to a much larger distance than those that can be
studied using stellar dynamics.

\acknowledgments

We thank Mike Eracleous for his help with observations and reduction of
ground-based spectra.
Support for Proposal \#9177 was provided by NASA through a grant
from the Space Telescope Science Institute, which is operated by
the Association of Universities for Research in Astronomy, Incorporated,
under NASA contract NAS5-26555.

\begin{deluxetable}{llccccccc}
\tabletypesize{\scriptsize}
\tablecaption{Summary of {\it ROSAT\/} Observations \label{latetbl1}}
\tablewidth{0pt}
\tablehead{
\colhead{Galaxy} &
\colhead{Date} &
\colhead{$\Gamma$} &
\colhead{L$_{X}$(pl)} &
\colhead{kT$_{\rm bb}$} &
\colhead{L$_{X}$(bb)} &
\colhead{N$_{H}$(Gal)} &
\colhead{t$_{\rm flare}$} &
\colhead{Variability} \\
\colhead{} &
\colhead{} & 
\colhead{($N(E)\propto E^{-\Gamma}$)} &
\colhead{(ergs s$^{-1}$)} &
\colhead{(eV)} &
\colhead{(ergs s$^{-1}$)} &
\colhead{(cm$^{-2}$)} &
\colhead{(months)} &
\colhead{Amplitude}
}
\startdata
RXJ1242.6--1119A & 1992 Jul & $5.1 \pm 0.9$ & $1.42 \times 10^{44}$ & $60 \pm 1$ & $3.5 \times 10^{43}$ & $3.74 \times 10^{20}$ & $<18$ & $>20$ \\
RXJ1624.9+7554 & 1990 Oct & $2.29 \pm 0.12$ & $1.1 \times 10^{44}$ & $97 \pm 4$ & $1 \times 10^{44}$  & $3.9 \times 10^{20}$ & $<18$ & $>200$ \\
NGC 5905 & 1990 Jul & $4.0 \pm 0.4$ & $6.1 \times 10^{41}$ & $56 \pm 12$ & $9.2 \times 10^{41}$ & $1.5 \times 10^{20}$ & $\sim 5$ & $>120$ \\
\enddata
\end{deluxetable}

\begin{deluxetable}{llllcccc}
\tabletypesize{\scriptsize}
\tablecaption{Log of Post-Flare Optical Spectra \label{latetbl2}}
\tablewidth{0pt}
\tablehead{
\colhead{Galaxy} &
\colhead{Date} &
\colhead{Telescope} &
\colhead{Grating or} &
\colhead{Slit Width} &
\colhead{Wavelength Range} &
\colhead{Resolution} &
\colhead{t$_{\rm exp}$} \\
\colhead{} &
\colhead{} & 
\colhead{} & 
\colhead{Spectrograph} &
\colhead{(arcsec)} &
\colhead{(\AA )} &
\colhead{(\AA )} &
\colhead{(min)}
}
\startdata
RXJ1242.6--1119A & 2001 Aug 9  & {\it HST\/}/STIS & G430L & 0.2 & 2900--5700 & 4.1 & 91.5 \\
               & 2001 Aug 9  & {\it HST\/}/STIS & G750L & 0.2 & 5240--10270 & 7.9 & 79.8 \\
               & 2000 Jun 4  & KPNO 2.1m      & Goldcam & 1.9 & 3900--7500 & 5   & 30.0 \\
               & 2000 Jun 1  & MDM 2.4m         & CCDS  & 1.5 & 3200--6800 & 12  & 30.0 \\ \\
RXJ1624.9+7554 & 2001 Oct 1  & {\it HST\/}/STIS & G430L & 0.2 & 2900--5700 & 4.1 & 101.5 \\
               & 2001 Oct 1  & {\it HST\/}/STIS & G750L & 0.2 & 5240--10270 & 7.9 & 92.2 \\
               & 2000 Jun 4  & KPNO 2.1m      & Goldcam & 1.9 & 3900--7500 & 5   & 30.0 \\
               & 1999 Feb 9  & MDM 2.4m         & MkIII & 1.7 & 4000--8600 & 5   & 40.0 \\ \\
NGC 5905       & 2001 Dec 20 & {\it HST\/}/STIS & G430L & 0.1 & 2900--5700 & 3.8 & 37.2 \\
               & 2001 Dec 20 & {\it HST\/}/STIS & G750M & 0.1 & 6295--6867 & 0.8 & 49.7 \\
               & 2000 Jun 4  & KPNO 2.1m      & Goldcam & 1.9 & 3900--7500 & 5   & 15.0 \\
               & 2000 Jun 1  & MDM 2.4m         & CCDS  & 1.5 & 3200--6800 & 12  & 15.0 \\
\enddata
\end{deluxetable}

\begin{deluxetable}{lccc}
\tabletypesize{\scriptsize}
\tablecaption{NGC 5905 Post-flare Optical Line Fluxes \label{latetbl3}}
\tablewidth{0pt}
\tablehead{
\colhead{Line} &
\colhead{KPNO 2.1m} &
\colhead{{\it HST\/}/STIS} &
\colhead{Dereddened}
}
\startdata
 [O II] $\lambda$3727  & . . .    & $0.19 \pm 0.01$ & $3.20 \pm 0.24$ \cr
 He~II $\lambda$4686   & $< 0.03$   & $< 0.02$ & $< 0.11$ \cr
 H$\beta\ \lambda$4861 & 0.27 & $0.08 \pm 0.01$ & $0.32 \pm 0.03$ \cr
 [O III] $\lambda$4959 & 0.05 & $0.22 \pm 0.02$ & $0.75 \pm 0.06$ \cr
 [O III] $\lambda$5007 & 0.12 & $0.58 \pm 0.03$ & $1.88 \pm 0.09$ \cr
 [O I] $\lambda$6300 & $< 0.03$ & $0.16 \pm 0.04$ & $0.19 \pm 0.05$ \cr
 [N II]  $\lambda$6548 & 0.15 & $0.53 \pm 0.05$ & $0.53 \pm 0.05$ \cr
 H$\alpha\ \lambda$6563 & 1.00\tablenotemark{a} & $1.00\pm 0.06$\tablenotemark{b} & $1.00 \pm 0.06$\tablenotemark{c} \cr
 [N II]  $\lambda$6583 & 0.44 & $1.53 \pm 0.07$ & $1.51 \pm 0.07$ \cr
 [SII] $\lambda$6716   & 0.14 & $0.11 \pm 0.04$ & $0.10 \pm 0.04$ \cr
 [SII] $\lambda$6731   & 0.11 & $0.18 \pm 0.03$ & $0.16 \pm 0.03$ \cr
\enddata
\tablenotetext{a}{F(H$\alpha$) = 3.7 x 10$^{-14}$ ergs cm$^{-2}$ s$^{-1}$.}
\tablenotetext{b}{F(H$\alpha$) = 1.0 x 10$^{-15}$ ergs cm$^{-2}$ s$^{-1}$.}
\tablenotetext{c}{F(H$\alpha$) = 2.4 x 10$^{-14}$ ergs cm$^{-2}$ s$^{-1}$.}
\end{deluxetable}

\begin{figure}[tbp] \label{specfig1} \figurenum{1}
\begin{center}
\plotone{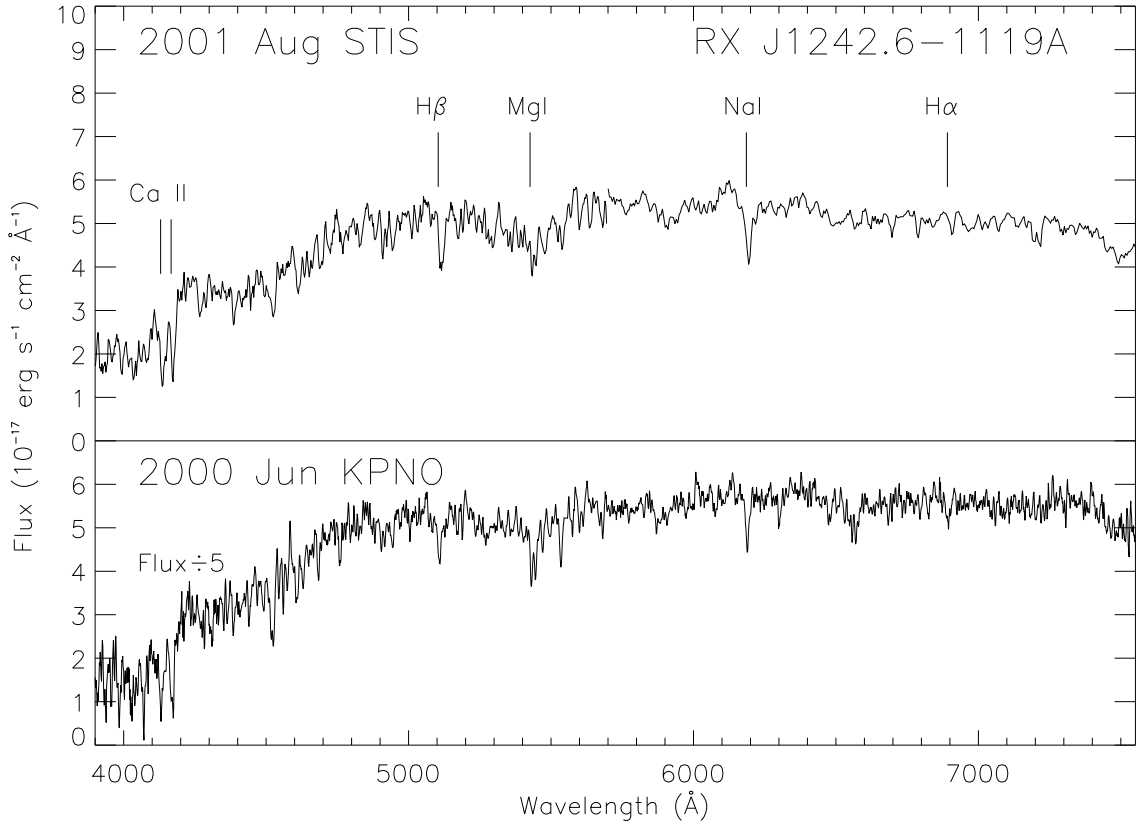}
\caption{Post-flare spectra of RX J1242.6--1119A from {\it HST\/}/STIS and the KPNO 2.1m,
smoothed over one spectral resolution element.  The KPNO spectrum has been divided
by a factor of 5 for comparison. 
Both spectra show only stellar absorption features, with no evidence of
high excitation emission lines or a non-stellar continuum indicative of an AGN.}
\end{center}
\end{figure}

\begin{figure}[tbp] \label{specfig2} \figurenum{2}
\begin{center}
\plotone{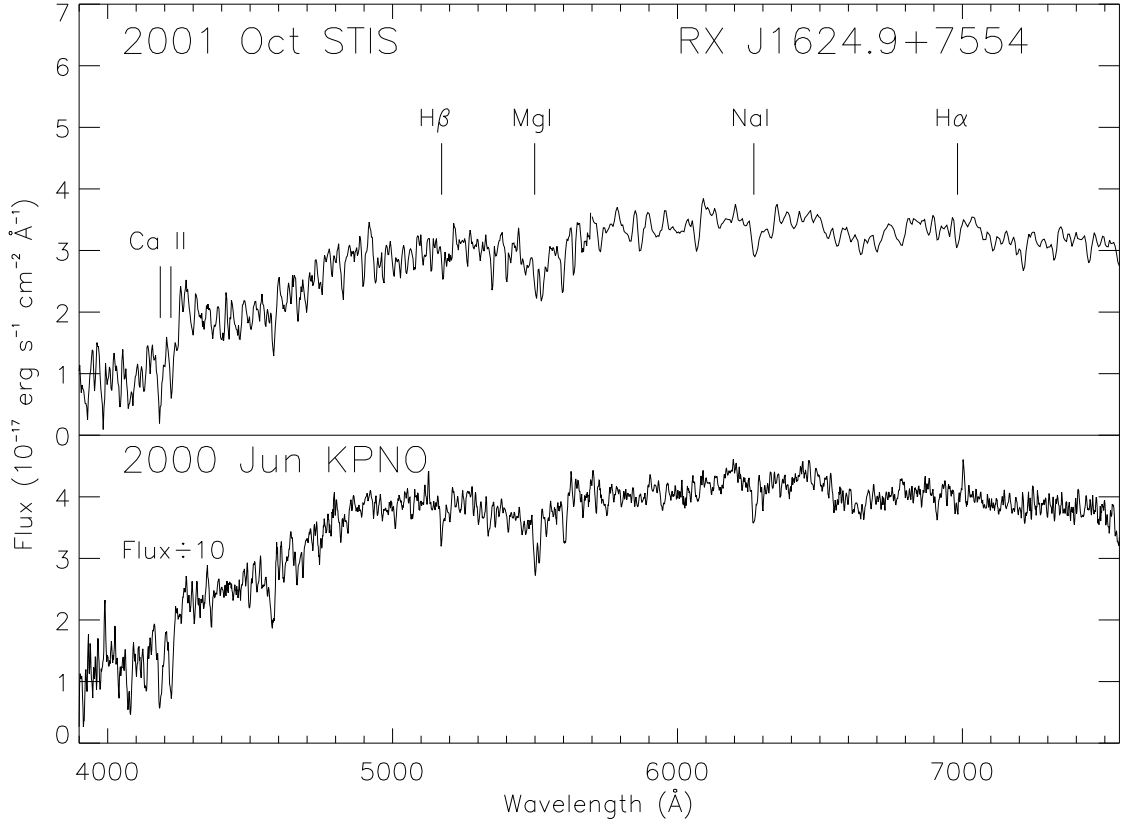}
\caption{Post-flare spectra of RX J1624.9+7554 from {\it HST\/}/STIS and the KPNO 2.1m,
smoothed over one spectral resolution element.  The KPNO spectrum has been divided by a 
factor of 10 for comparison.  Both spectra show similar stellar
absorption features, with no evidence of high-excitation emission lines or a non-stellar 
continuum indicative of an AGN.  Weak, extended [N II]$\lambda$6583
emission is detected in the KPNO spectrum, 
most likely associated with H~II regions near the nucleus.}
\end{center}
\end{figure}

\begin{figure}[tbp] \label{statsfig3} \figurenum{3}
\begin{center}
\plotone{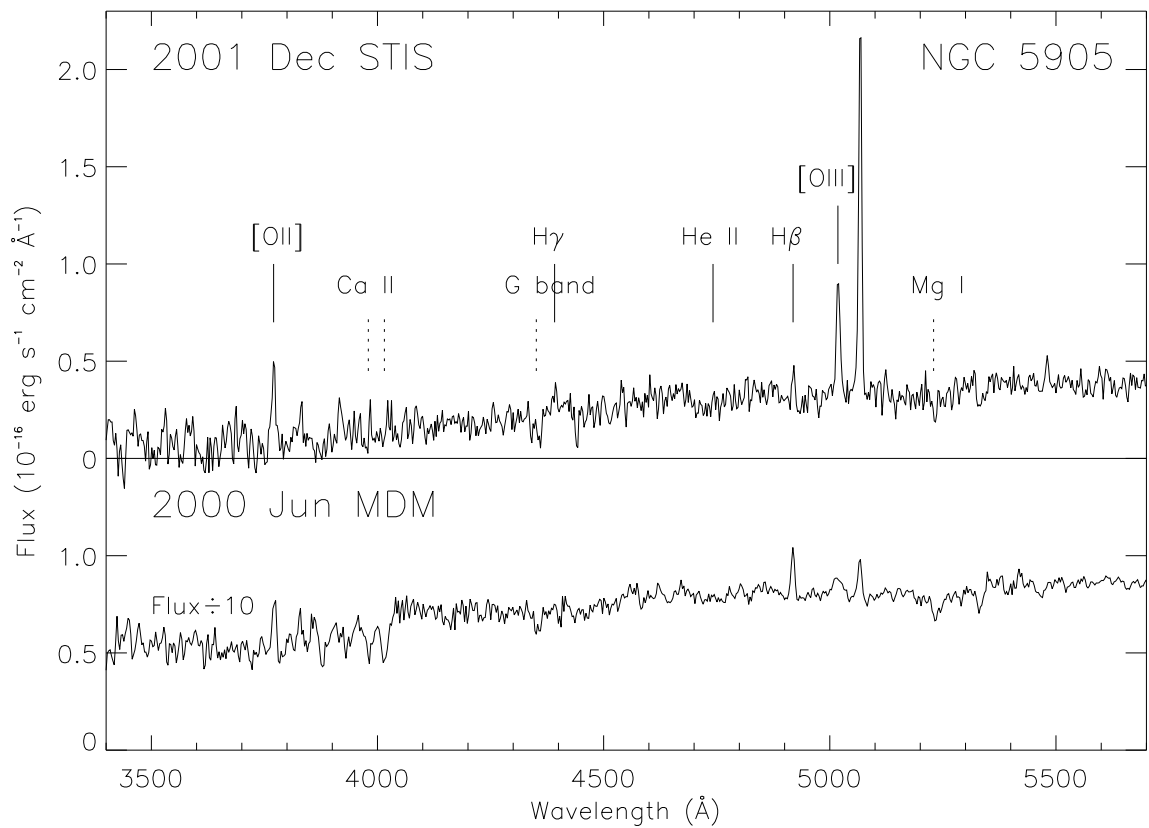}
\caption{Post-flare spectra of the nucleus of NGC 5905
from {\it HST\/}/STIS and the MDM 2.4m,
smoothed over one spectral resolution element.  The MDM spectrum 
has been divided by 10 for comparison.  
The STIS spectrum has narrow-line emission consistent with a Seyfert
2 nucleus, in contrast to the MDM spectrum which is classified as an H~II/starburst.
Stellar absorption features, shown with dotted lines, are
observed in both the STIS and ground-based nuclear spectra,
indicating the lack of a strong non-stellar continuum.}
\end{center}
\end{figure}

\begin{figure}[tbp] \label{statsfig4} \figurenum{4}
\begin{center}
\plotone{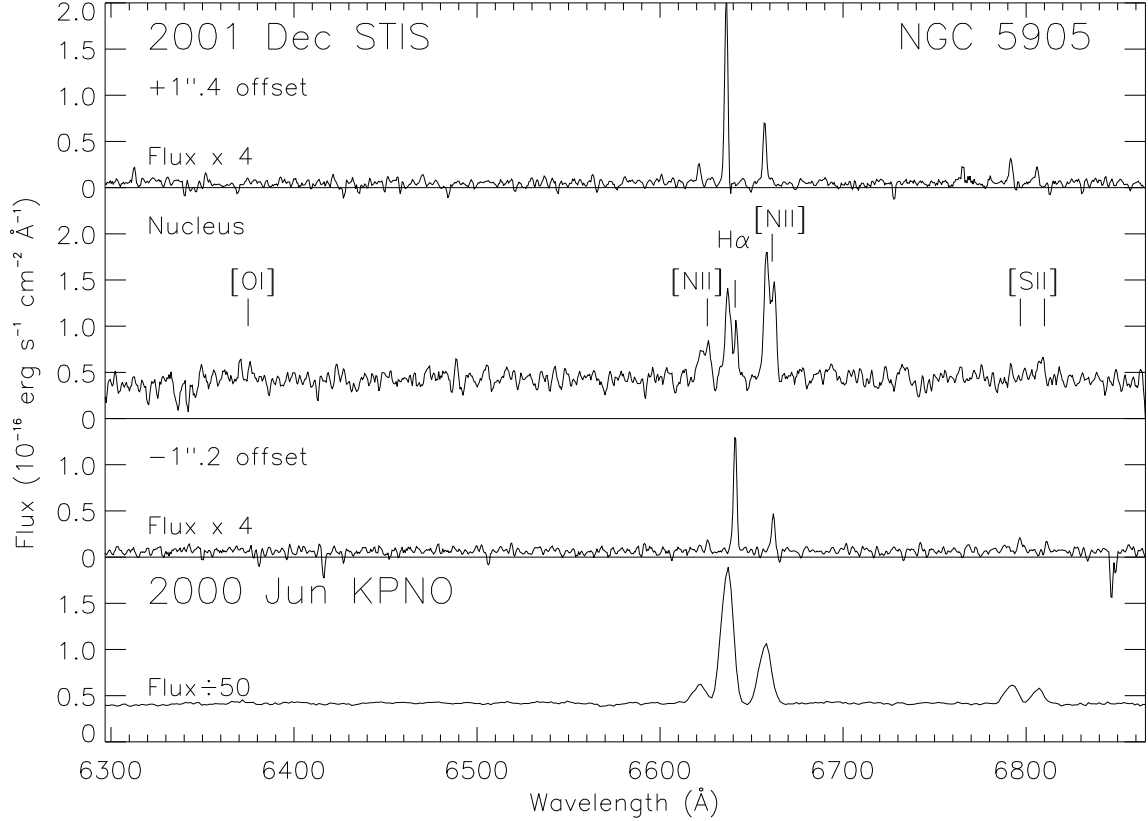}
\caption{Post-flare {\it HST\/}/STIS spectra of the nucleus of NGC 5905 with
the G750M grating, as well as regions of peak H~II emission offset $+1.\!^{\prime\prime}4$ and $-1.\!^{\prime\prime}2$ from the nucleus, smoothed over one spectral resolution element.
The off-nuclear STIS spectra have been multiplied by 4 for comparison.
A spectrum from the KPNO 2.1m, divided by 50, is shown for comparison.
The STIS spectrum has narrow-line emission consistent with
a Seyfert 2 nucleus, while the off-nuclear STIS spectra and ground-based spectrum 
have narrow-line emission consistent with an H~II/starburst. 
The double peaks of H$\alpha$ and [N II] emission in the nuclear STIS spectrum match the
velocity shifts of the off-nuclear spectra, which can be ascribed to the
rotation curve (see Figure 5).}
\end{center}
\end{figure}

\begin{figure}[tbp] \label{statsfig5} \figurenum{5}
\begin{center}
\plotone{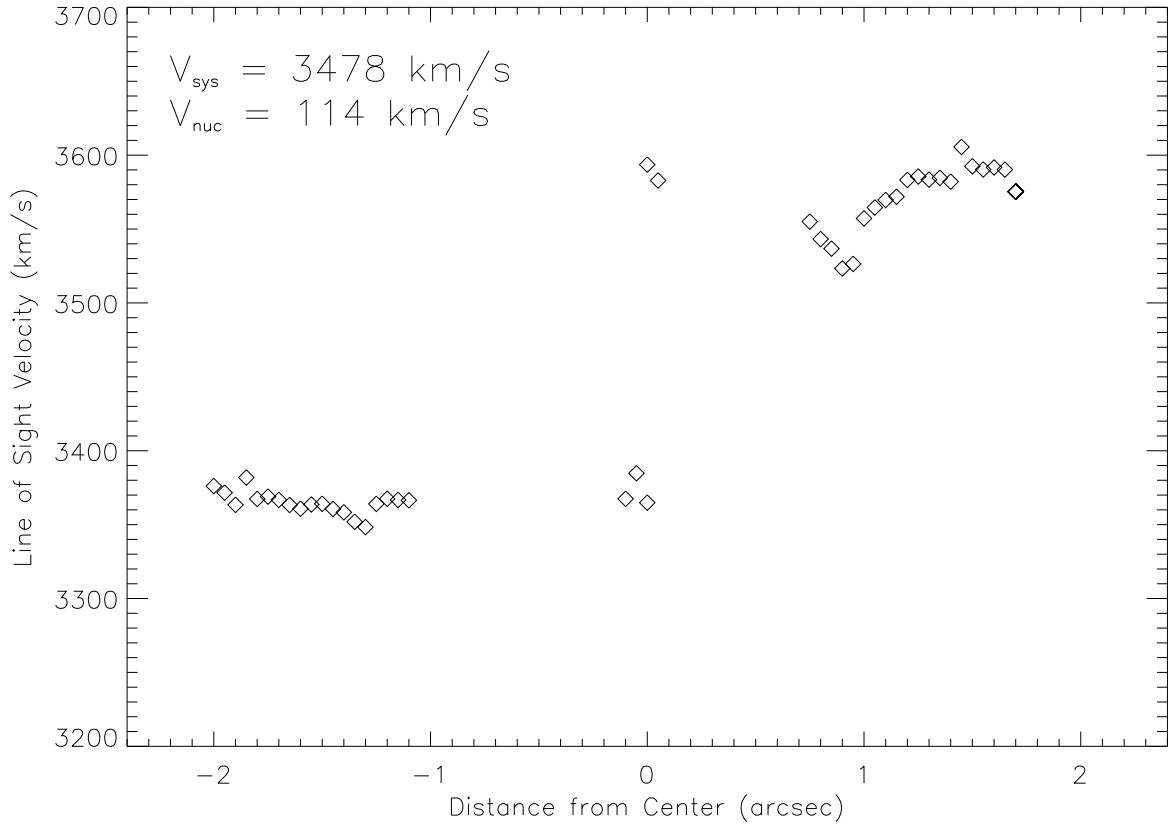}
\caption{H$\alpha$ rotation curve of NGC 5905 measured from
$0.\!^{\prime\prime}05$ (1 pixel) wide rows on the STIS CCD. 
%with dotted line showing the systematic velocity, and nuclear velocity.
%Error bars show $1\sigma$ error.
}
\end{center}
\end{figure}

\begin{figure}[tbp] \label{specfig6} \figurenum{6}
\begin{center}
\plotone{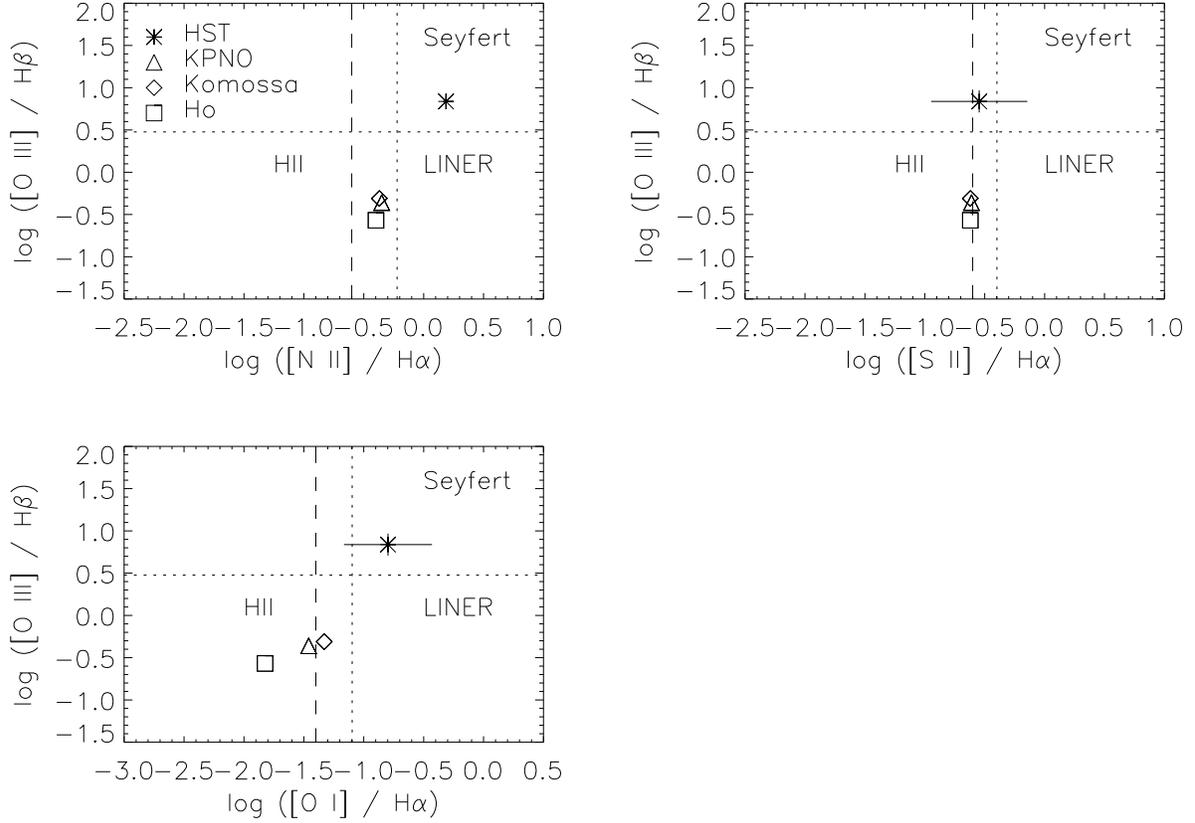}
\caption{Diagnostic diagrams of Veilleux \& Osterbrock (1987), with {\it HST\/}
and various ground-based measurements of emission lines in NGC 5905.
The {\it HST\/} points are shown with $3\sigma$ error bars.
The {\it short dashed\/} lines are the boundaries separating different
classes of objects from Veilleux \& Osterbrock (1987), while the
{\it long dashed\/} line is the classification boundary from Ho et al. (1997).}
\end{center}
\end{figure}

\begin{figure}[tbp] \label{specfig7} \figurenum{7}
\begin{center}
\plotone{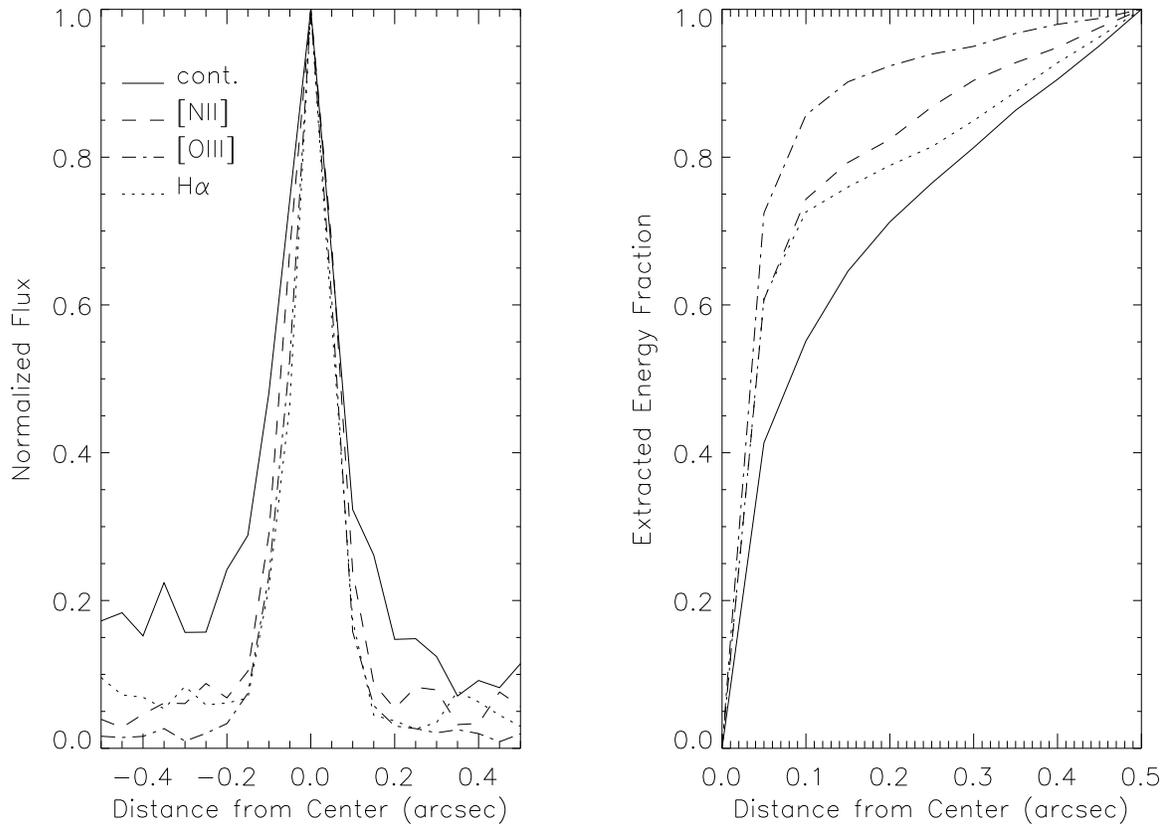}
\caption{Spatial profile of the 2D spectrum of NGC 5905
for [O III], [N II], and H$\alpha$ emission lines
perpendicular to the dispersion direction, as well as the profile
of the mean continuum from $5480-5530$ \AA, showing that the starlight
is more extended than the nuclear emission lines.}
\end{center}
\end{figure}

\begin{figure}[tbp] \label{specfig8} \figurenum{8}
\begin{center}
\plotone{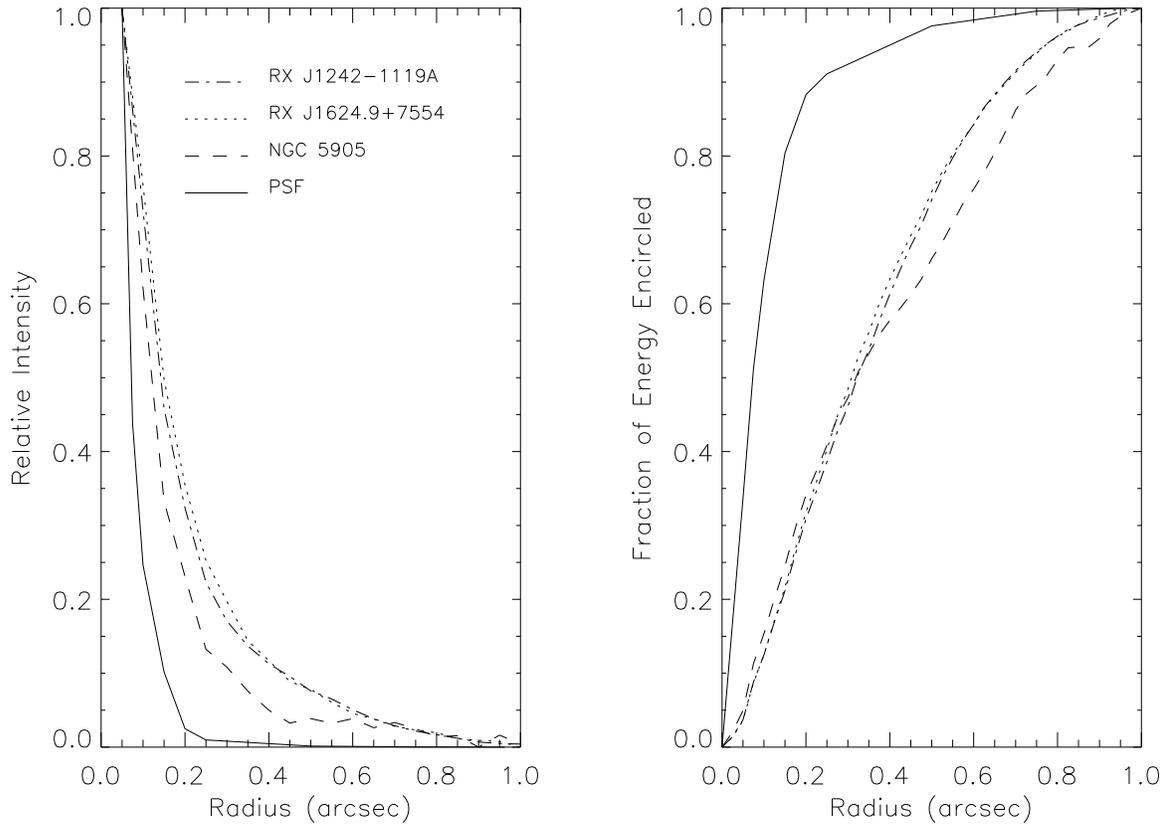}
\caption{Radial profiles of the acquisition image of the three targets,
with a background level measured outside of $1^{\prime\prime}$ subtracted,
in comparison with the PSF.}
\end{center}
\end{figure}

\end{document}